\documentclass{emulateapj}
\usepackage{epsfig}
\usepackage{color}
\newcommand{\msun}{\,\rm M_\odot}

\newcommand{\kmsmpc}{\,{\rm km\,s^{-1}\,Mpc^{-1}}}
\newcommand{\be}{\begin{equation}}
\newcommand{\ee}{\end{equation}}
\newcommand{\ba}{\begin{eqnarray}}
\newcommand{\ea}{\end{eqnarray}}
\newcommand{\f}{\frac} 
\newcommand{\rvir}{r_{\rm vir}}
\newcommand{\mvir}{M_{\rm halo}}
\newcommand{\msub}{M_{\rm sub}}

\newcommand{\kms}{\,{\rm km\,s^{-1}}}

\lefthead{Diemand, Kuhlen, \& Madau}
\righthead{Dark matter substructure in the MW}



\begin{document}
\submitted{Submitted Aug 17, 2006}
\accepted{Nov 6, 2006}

\title{Dark matter substructure and gamma-ray annihilation in the Milky Way halo}
\author{J\"urg Diemand\altaffilmark{1,2}, Michael Kuhlen\altaffilmark{1,3},
\& Piero Madau\altaffilmark{1,4}}

\altaffiltext{1}{Department of Astronomy \& Astrophysics, University of
California, Santa Cruz, CA 95064.}
\altaffiltext{2}{Hubble Fellow.}
\altaffiltext{3}{School of Natural Sciences, Institute for Advanced Study,
Einstein Drive, Princeton, NJ 08540.}
\altaffiltext{4}{Max-Planck-Institut fuer Astrophysik, Karl-Schwarzschild-Strasse
1, 85740 Garching bei Muenchen, Germany.}
\email{diemand@ucolick.org}

\begin{abstract}
We present initial results from ``Via Lactea'', the highest resolution simulation 
to date of Galactic CDM substructure. It follows the formation
of a Milky Way-size halo with $\mvir=1.8\times 10^{12}\,\msun$ in 
a {\it WMAP} 3-year cosmology, using 234 million particles. 
Over 10,000 subhalos can be identified at z=0:
Their cumulative mass function is well-fit by $N(>\msub)=
0.0064\,(\msub/\mvir)^{-1}$ down to $\msub=4\times 10^6\,\msun$. The total mass fraction 
in subhalos is 5.3\%, while the fraction of surface mass density in 
substructure within a projected distance of 10 kpc from the halo center
is 0.3\%. Because of the significant contribution from the smallest resolved subhalos,
these fractions have not converged yet. Sub-substructure is apparent in all the larger satellites,
and a few dark matter lumps are resolved even in the solar vicinity. 
The number of dark satellites with peak circular velocities 
above $10\,\kms$ ($5\,\kms$) is 124 (812): of these, 5 (26) are found within $0.1\,\rvir$, a region 
that appeared practically smooth in previous simulations.
The neutralino self-annihilation $\gamma$-ray emission from dark 
matter clumps is approximately constant per subhalo mass decade. Therefore,
while in our run the contribution of substructure to the $\gamma$-ray 
luminosity of the Galactic halo amounts to only $40$\% of the total spherically-averaged 
smooth signal, we expect this fraction to grow significantly as resolution is increased
further. An all-sky map of the expected annihilation $\gamma$-ray flux reaching a
fiducial observer at 8 kpc from the Galactic center shows that at the current
resolution a small number of subhalos start to be bright enough to
be visible against the background from the smooth density field
surrounding the observer.
\end{abstract}

 
\section{Introduction}

In structure formation scenarios dominated by cold dark matter (CDM), the halos 
of galaxies and galaxy clusters are assembled via the hierarchical merging and 
accretion of smaller progenitors. This process causes structures to virialize to a 
new equilibrium by redistributing energy among the collisionless 
mass components. Early low resolution numerical simulations and simple analytical models found
that the end products of this ``bottom-up'' scenario were smooth, triaxial halos. In recent 
years, higher resolution cosmological N-body simulations have modified this picture: the merging 
of progenitors is not always complete, and the cores of accreted halos often survive as 
gravitationally bound subhalos orbiting within a larger host system. CDM halos are
not smooth, they have a wealth of substructure on all resolved mass
scales (e.g. \citealt{Moore1999,Klypin1999,Moore2001,Stoehr2003,Diemand2004sub,Reed2005,Gao2005,Diemand2006}). 

The amount and spatial distribution of subhalos around their host
provide unique information and clues on the galaxy assembly process and
the nature of the dark matter. The mismatch between the dozen or
so dwarf satellite galaxies observed around the Milky Way and the predicted large 
number of CDM subhalos    
of comparable circular velocity \citep{Moore1999,Klypin1999} has been the 
subject of many recent studies. Solutions
involving feedback mechanisms that make star formation in small halos very inefficient 
offer a possible way out (e.g. \citealt{Bullock2000,Somerville2002,Kravtsov2004,Moore2006}). 
Other models have attempted to solve the apparent small-scale problems of CDM at a more
fundamental level, i.e. by reducing small-scale power (e.g. 
\citealt{Kamionkowski2000}). Even if most dark matter satellites have no 
optically luminous counterparts, the substructure population 
may be detectable via flux ratio anomalies in strong gravitational lenses 
(e.g. \citealt{Metcalf2001,Chiba2002}), through its effects on stellar 
streams (e.g. \citealt{Ibata2002,Mayer2002}), or possibly via $\gamma$-rays 
from dark matter annihilation in their cores \citep{Bergstrom1999,CalcaneoRoldan2000,
Stoehr2003,Colafrancesco2005,Diemand2006}. 

The possibility of observing the fingerprints of the small-scale
structure of CDM hinges on the ability of subhalos to survive the
hierarchical clustering process as substructure within the host. This
is in turn particularly sensitive to resolution issues, as subhalos
with numerically softened central densities are more easily disrupted
by tidal forces. In this paper we present initial results on halo
substructure from a new N-body cosmological simulation of
unprecedented dynamic range: dubbed ``Via Lactea'', it resolves the galaxy forming
region of a Milky Way-size halo at $z=0$ with over 200 million particles. 
This is an order of magnitude more than used in previous
simulations. The run was completed in 320,000 CPU hours on NASA's
Project Columbia supercomputer, currently one of the fastest machines
available.
\begin{figure*}
\plotone{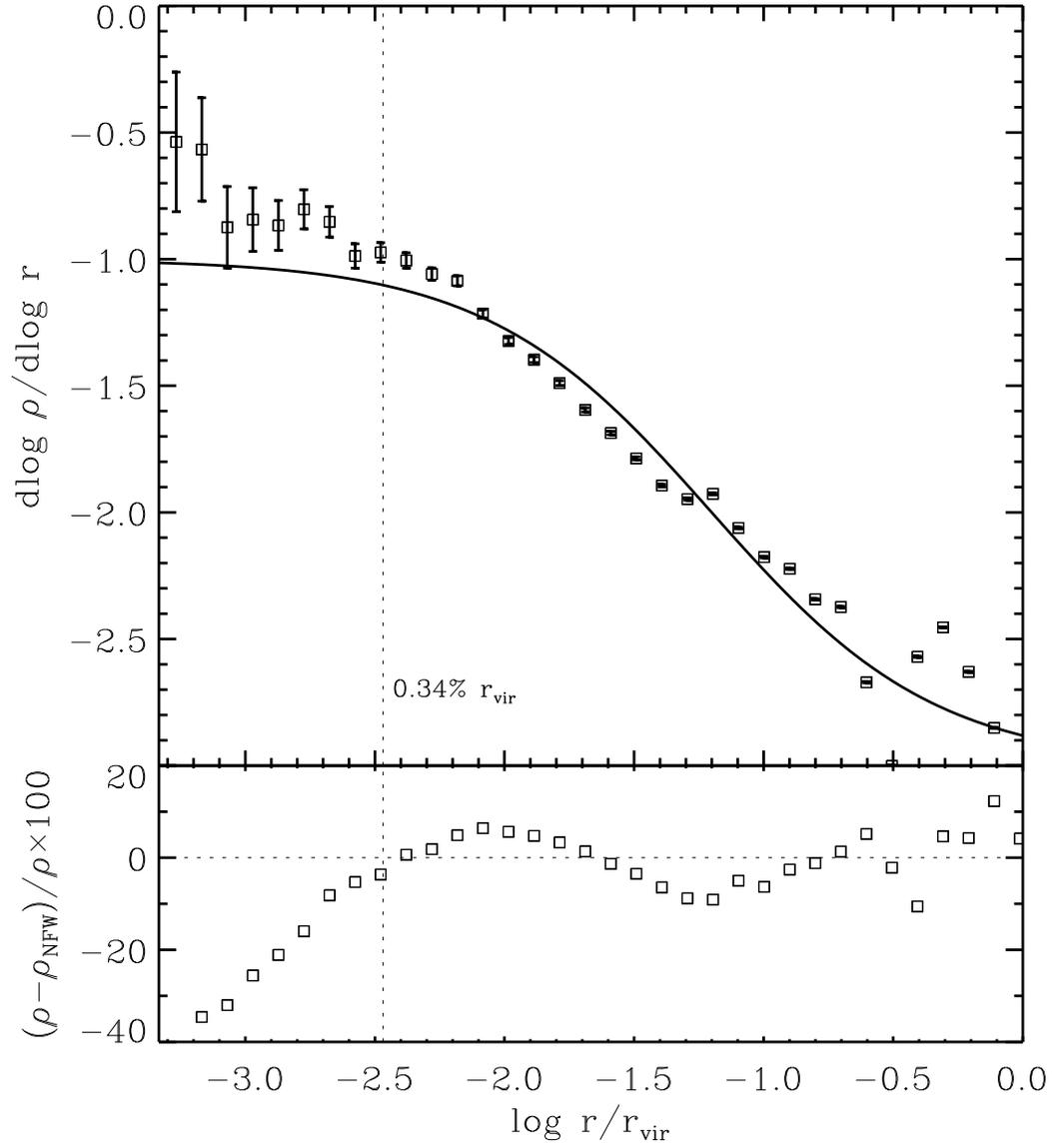}
\caption{\textit{Top}: logarithmic slope of the density profile of our Via Lactea
run, as a function of radius. Densities were computed in 50 radial
logarithmic bins, and the local slope was determined by a finite
difference approximation using one neighboring bin on each side. The
thin line shows the slope of the best-fit NFW profile with
scale radius of $24.6$ kpc. The vertical dotted line indicates the
estimated convergence radius: local densities 
(but not necessarily the logarithmic slopes) 
should be correct to within 10\% outside of this radius.
\textit{Bottom}: the residuals in
percent between the density profile and the best-fit NFW profile, as a
function of radius.
}
\vspace*{0.1in}
\label{profile}
\end{figure*}

\section{Simulations}

The simulation was performed with PKDGRAV
\citep{Stadel2001,Wadsley2004} and employed multiple mass particle
grid initial conditions generated with the GRAFICS package
\citep{Bertschinger2001}. The high resolution region was sampled with
234 million particles particles of mass $2.1\times 10^4\msun$ and evolved
with a force resolution of 90 pc. It was embedded within a periodic
box of comoving size 90 Mpc, which was sampled at lower resolution to
account for the large scale tidal forces. We adopted the best-fit
cosmological parameters from the {\it WMAP} three-year data release
\citep{Spergel2006}: $\Omega_M = 0.238$, $\Omega_\Lambda = 0.762$,
$H_0= 73\,\kmsmpc$, $n=0.951$, and $\sigma_8=0.74$. The lower values
of $\sigma_8$ and $n$ compared to 1-year {\it WMAP} results have the
effect of delaying structure formation and reducing small-scale power.
The simulation was centered on an isolated halo that had no major
merger after $z=1.7$, making it a suitable host for a Milky Way-like
disk galaxy. In this work we focus on results obtained at z=0, 
the formation history of the Via Lactea halo and
its substructures will be presented in a subsequent paper.
The host halo mass at $z=0$ is $\mvir=1.77\times
10^{12}\,\msun$ within a radius of $\rvir=389\,$ kpc (We define "$\rvir$"
as the radius within which the enclosed average density is 200
times the mean matter value). We used adaptive time-steps as short as
$(13.7/2\times 10^5)\,$Gyr $=68,500$ yr satisfying the condition
$\Delta t < 0.2 \sqrt{\epsilon/a}$, where $\epsilon$ is the force
softening length and $a$ is the local acceleration. This criterion is
insufficient to ensure convergence at small radii, where the dynamical time becomes too short 
and runs using the $\sqrt{\epsilon/a}$ criterion start to
underestimate matter densities. In comparable runs this problem was observed to 
to give 10\% too low local densities at
$6\times 10^5 \, \rho_{\rm crit}(z=0)$, see Figure 1 of \citealt{Diemand2005cusps}.
This density is reached at $0.0034\,\rvir$ in the Via Lactea run, therefore we set
$r_{\rm conv}=0.0034\,\rvir$. The local densities (but not necessarily the logarithmic slopes) 
should be correct to within 10\% beyond $0.0034\,\rvir$, but detailed convergence
tests using galaxy scale halos would be needed to confirm this. 
To resolve the density profile further in than $0.0034\,\rvir$, adaptive time-steps 
that scale with dynamical timescale should be adopted instead \citep{Zemp2006}.  The
force calculations in the tree algorithm PKDGRAV includes multipole
moments up to hexa-decapole to reach high force accuracy.
To check for numerical convergence of substructure properties we resimulated the same halo 
(plus a few neighboring galaxy halos) with 
27 times more massive particles. The parameters of the two simulations are summarized
in Table~\ref{tab:parameters}.
\begin{table}
\begin{center}
\begin{tabular}{ccccc}
\hline
\hline
$z_i$ & $z_f$ & $\epsilon$     & $N_p$ & $m_p$ \\
\hline
48.4 & 0.0 & 90 pc& $2.34 \times 10^8$ & $2.09\times 10^{4} \msun$ \\
48.4 & 0.0 & 378 pc & $6.50 \times 10^7$ & $5.64\times 10^{5} \msun$ \\
\end{tabular}
\end{center}
\tablecomments{Initial and final redshifts $z_i$ and $z_f$, (spline)
softening length $\epsilon$, total number $N_p$ and mass $m_p$ of dark matter
particles for the Via Lactea simulation ({\it top}) and a lower resolution run 
({\it bottom}). Force softening lengths $\epsilon$ are constant
in physical units back to $z=9$ and constant in comoving units before.}
\label{tab:parameters}
\end{table}

Figure \ref{profile} (upper panel) shows the logarithmic slope of the spherically-averaged 
density profile, $d\log\rho/d\log r$, for our Via Lactea halo, plotted versus radius. 
The fitting formula proposed by \citet*{NFW97} with scale radius of $24.6$ kpc
provides a reasonable approximation to the density profile down to 
our convergence radius $r_{\rm conv}$.  Within the region of convergence, deviations from 
the best-fit NFW matter density are typically less than 10\%. From $0.025\,\rvir$ down to 
$r_{\rm conv}$ Via Lactea is actually denser than predicted by the NFW formula. 
Near $r_{\rm conv}$ the density approaches the NFW value again while the logarithmic 
slope is shallower ($-1.0$ at $r_{\rm conv}$) than that the NFW fit. 
While we have checked that these conclusions are robust to 
variations in the binning used to construct the density profile,
they might be affected by numerical flattening near 
our convergence radius. We plan to run new simulations
with improved time-stepping \citep{Zemp2006} to address this issue.
\begin{figure*}
\centerline{\epsfig{figure=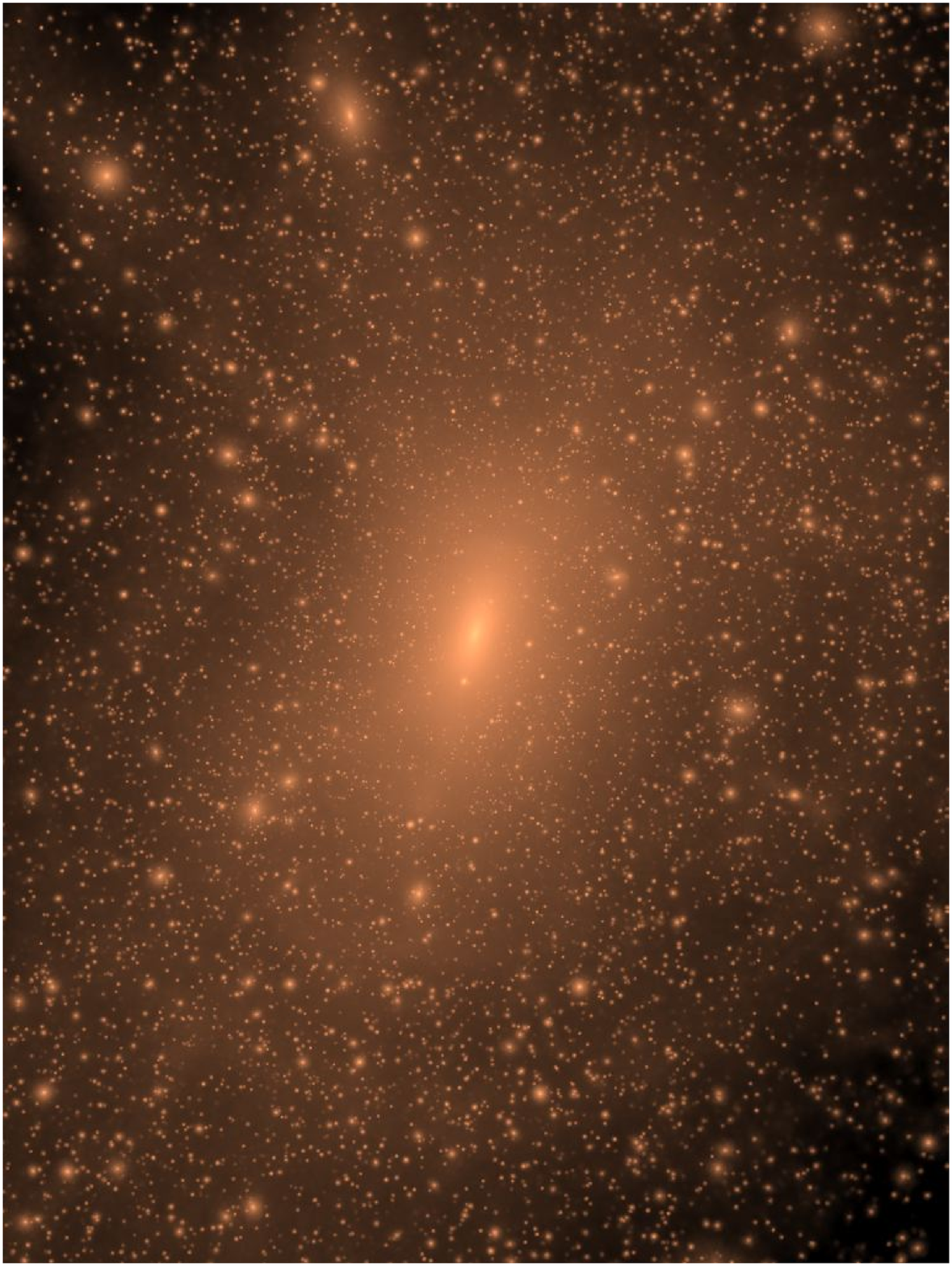,width=7in}}
\vspace{0.5cm}
\caption{Projected dark matter density-squared map of our simulated
Milky Way-size halo (``Via Lactea'') at the present epoch. The image covers
an area of 800 $\times$ 600 kpc, and the projection goes through a 600
kpc-deep cuboid containing a total of 110 million particles. The
logarithmic color scale covers 20 decades in density-square.  }
\label{L200}
\end{figure*}
\begin{figure*}
\centerline{\epsfig{figure=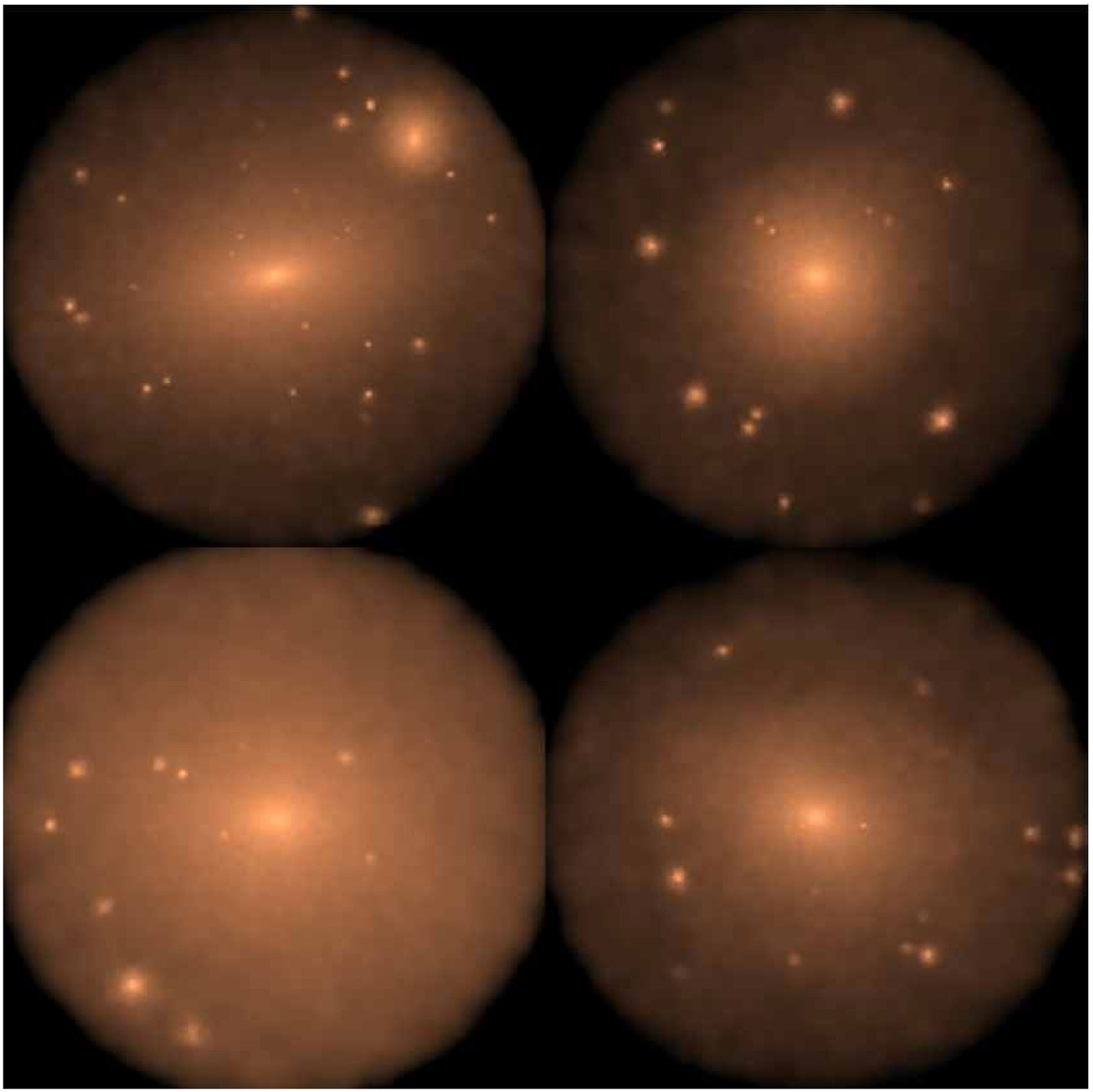,width=7in}}
\vspace{0.5cm}
\caption{Projected dark matter density-square map of the four most massive
subhalos within the simulated Milky Way host at the present epoch. Sub-substructure
is clearly visible. Only dark matter particles within the tidal radius $r_t$ 
are used for the projections. {\it Clockwise from top left}: ($\msub, r_t, r_{V_{\rm max}})=
(9.8\times10^9\,\msun, 40.1\,{\rm  kpc}, 7.6\,{\rm  kpc}),
(3.7\times10^9\,\msun, 33.4\,{\rm  kpc}, 4.0\,{\rm  kpc}), 
(3.0\times10^9\,\msun, 28.0\,{\rm  kpc}, 4.9\,{\rm  kpc})$, and 
($2.4\times10^9\,\msun, 14.7\,{\rm kpc}, 6.1\,{\rm  kpc})$.
The mean subhalo densities within the tidal radius (in units of the cosmic background dark matter
density) are 1002, 654, 904, and 4950, respectively. These values are related to 
the local matter density of the host (72, 46, 59 and 397 in the same units), and correlate only
weakly with the subhalo distance from the Galactic center (345, 374, 280 and 185 kpc).
}
\label{quad}
\end{figure*}
\section{Substructure in galaxy halos}

The wealth of substructure resolved in our Via Lactea run is clearly see in Figure \ref{L200}.
About 10,000 surviving satellites, some sub-subhalos 
(a few examples are shown in Fig. \ref{quad}), and 
even a couple of dark matter clumps at less than 8 kpc from the Galactic Center are 
now resolved.

\subsection{Subhalo identification}

\begin{figure*}
\centerline{\epsfig{figure=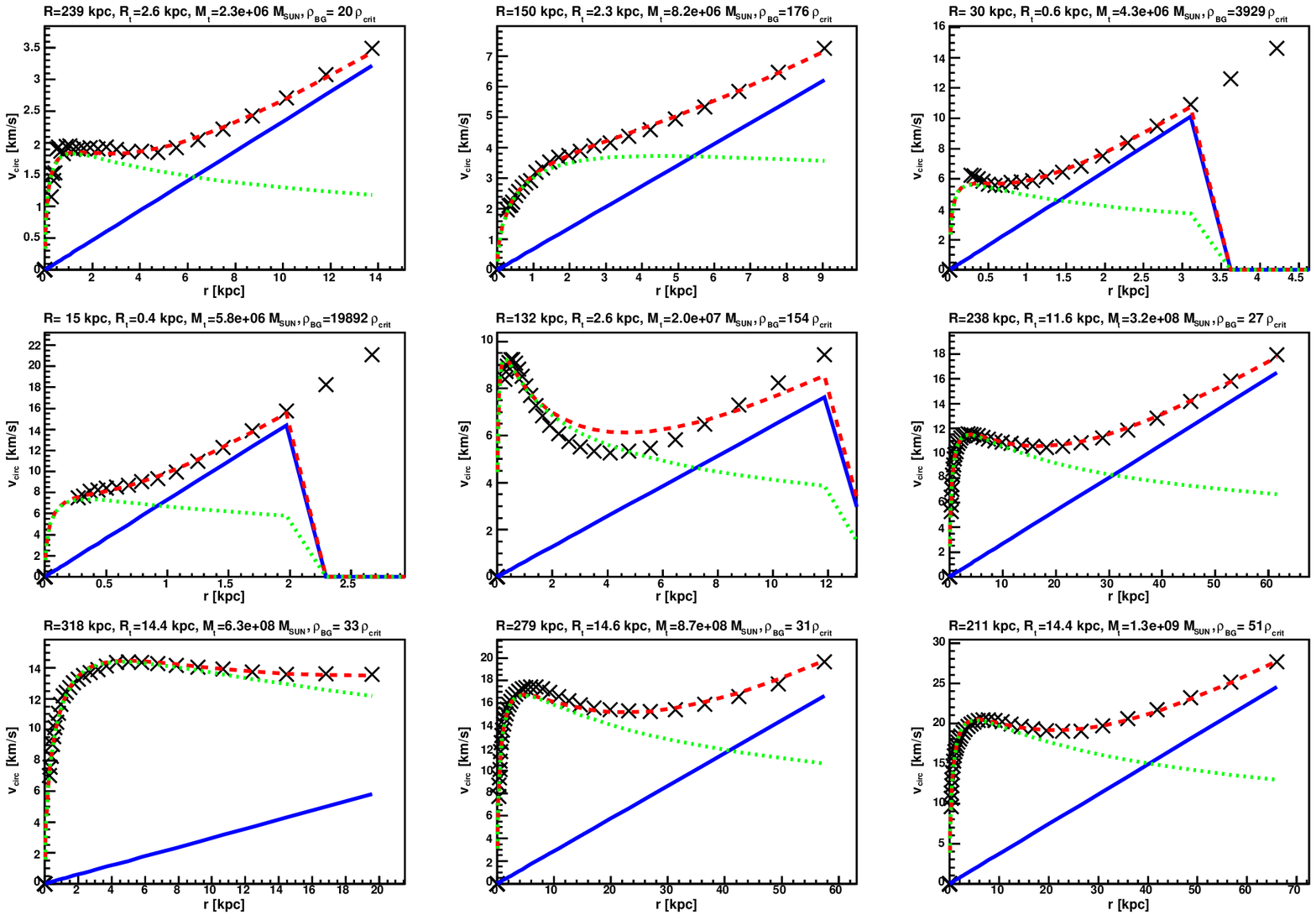,width=7in}}
\vspace{0.5cm}
\caption{Examples of subhalo circular velocity profiles ({\it crosses}).
The fitting functions ({\it dashed lines}) are
combinations of an NFW subhalo ({\it dotted lines}) and a linear contribution
form a constant background density ({\it solid lines}). Stated above each panel 
are the subhalos distance form the galaxy center, tidal radius, tidal mass and the
local background density.}
\vspace*{0.1in}
\label{tidal}
\end{figure*}

We have constructed substructure catalogs using the phase-space friends-of-friends 
algorithm 6DFOF described in \citet{Diemand2006}. We adopt a space linking 
length of 0.28 physical kpc (0.033 of the mean particle separation at $z=0$), 
a velocity space linking length of 16 physical $\kms$ (0.1 of the mean 1D velocity
dispersion of the host halo), and keep all groups with at least 16 members. 
Around their centers, spherical density profiles (using 50 logarithmic bins) were 
constructed: the bins reach out to $\rvir$ in the case of isolated halos or out 
to the center of the nearest larger halo in the case of subhalos, i.e. all 
structures are allowed to contain their own, smaller substructures. The resulting circular 
velocity profiles were then fitted with the sum of an NFW profile
and a constant density background. The radial
range used in these fits starts at 3 force softening lengths, 
i. e. 0.27 kpc in the Via Lactea run, and ends either 
at the largest radial bin stored during at runtime or further in if the circular
velocity grows linearly (within 10\%) over five consecutive bins.
Figure \ref{tidal} shows nine example
subhalo circular velocity profile fits.
A subhalo tidal radius $r_{t}$ was 
defined as $\rho_{\rm sub}(r_t)=2\times \rho_{\rm BG}$, where $\rho_{\rm BG}$ is the 
local matter density of the host halo. This is the tidal radius of an isothermal 
($\rho_{\rm sub}\propto r^{-2}$) satellite
on a circular orbit within an isothermal host.  
We define the subhalo mass to be all the mass within $r_{t}$, i.e. we did not perform 
an unbinding procedure. The speed of 6DFOF and its
parallel implementation into PKDGRAV allowed us to use it at run-time
with negligible computational cost (compared to the gravity calculations) on
all 200 snapshots that were stored (separated by 68.5 Myr). 

The 6DFOF subhalo list was checked in
various regions of the Via Lactea halo against a list obtained with SKID 
\citep{Stadel2001}.
The results compare very
well down to about 40 particles per halo except for the very largest subhalos, 
where the SKID algorithm tends to underestimate the masses and
is unable to properly account for the presence of resolved sub-substructure.
Detailed evolutionary tracks of subhalo properties and density profiles 
will be published in a subsequent paper.

\begin{figure*}
\plotone{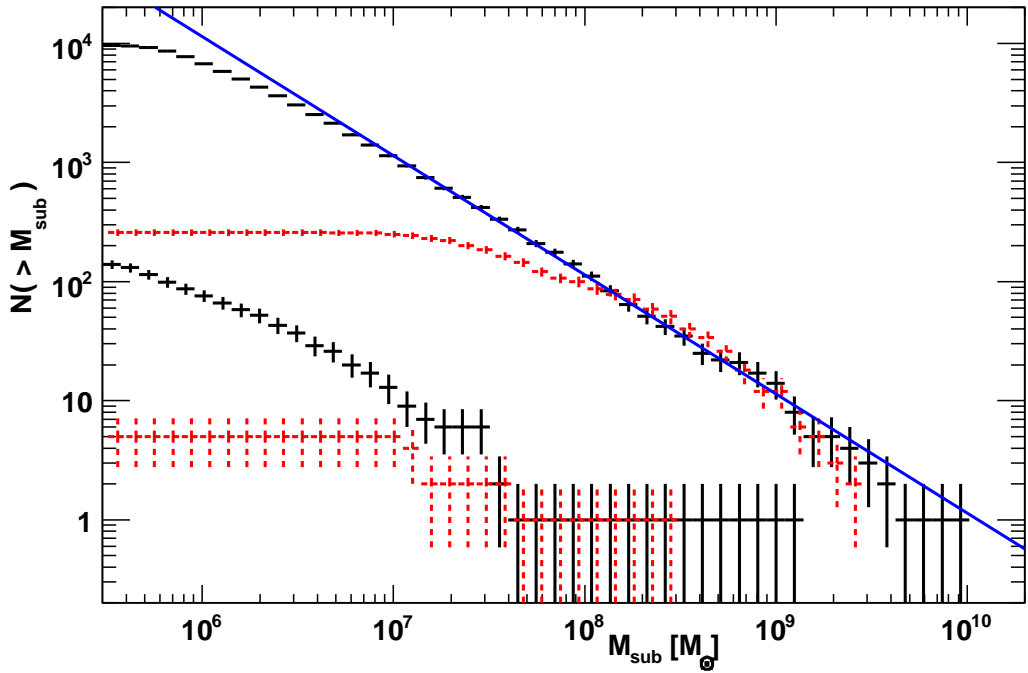}
\caption{Cumulative subhalo mass functions within $\rvir$ ({\it upper 
curves}) and $0.1\,\rvir$ ({\it lower curves}). {\it Solid crosses}: 
Via Lactea run. {\it Dashed crosses}: lower resolution run. {\it Solid line}:
power-law fit, $N(>\msub)=0.0064\times(\msub/\mvir)^{-1}$.
}
\vspace*{0.1in}
\label{massf}
\end{figure*}

\subsection{Subhalo abundance}

In Figure \ref{massf} we present the cumulative mass function for our
entire subhalo populationw ithin $\rvir$ as well as for the
subpopulation within the inner $0.1\,\rvir$. The former is well-fit by
a power-law $N(>M_{\rm sub})=0.0064 (\msub/\mvir)^{-1}$ above
$\msub=4\times 10^6\,\msun$, as shown by the solid line (formally, the
best-fit slope in the mass range $\msub=4\times 10^6-4\times
10^9\,\msun$ is $-0.97\pm0.03$).  This slope implies an equal mass per
subhalo mass decade, making the total mass in substructure quite
dependent on resolution. Note how numerical resolution effects start
to flatten the distribution already below masses corresponding to
about 200 particles (i.e. below $\msub=4\times10^6\, \msun$ and
$\msub=10^8\,\msun$ for the Via Lactea and the lower resolution run,
respectively).  The fraction of the host halo mass in subhalos is 
$f_{\rm sub}=5.3\%\, (0.87\%)$ within 
$\rvir\, (0.1\,\rvir)$. The contribution from subhalos below
$10^8\,\msun$ is significant: $f_{\rm
sub}(\msub<10^8\,\msun)=2.3\%\, (0.15\%)$.  The total substructure mass
fractions in the lower resolution run are substantially smaller,
$f_{\rm sub}=2.6\%\, (0.11\%)$. This is in part because 
the low resolution realization happens
to contain slightly fewer massive satellites at $z=0$, 
and in part because of the reduced number of resolved small subhalos.

\begin{figure*}
\plotone{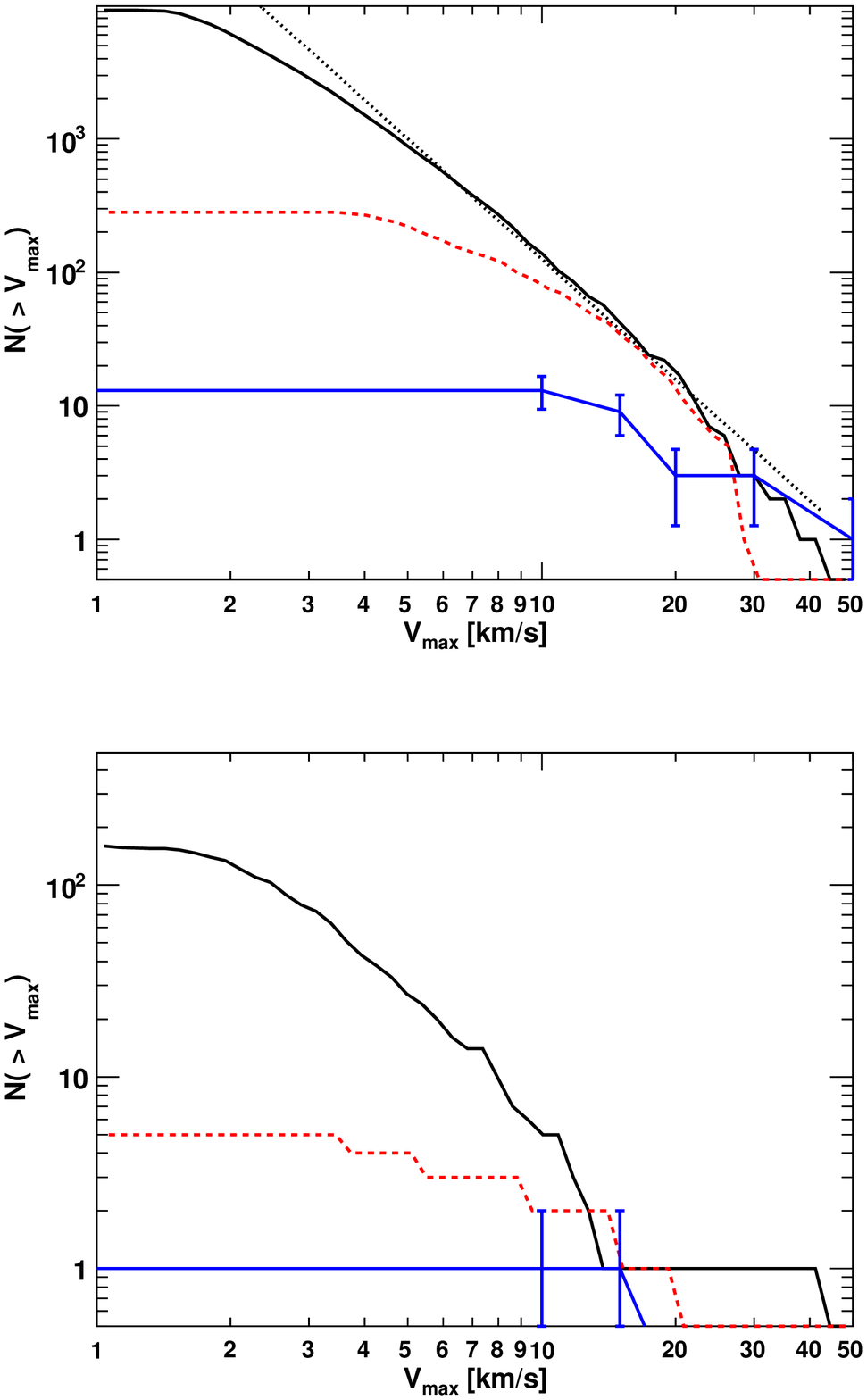}
\caption{Cumulative peak circular velocity function for all subhalos within 
$\rvir$ ({\it upper panel}) and for the subpopulation within the inner $0.1\,\rvir$ 
({\it lower panel}). {\it Solid lines without error bars}: 
Via Lactea run. {\it Dashed lines}: lower resolution run. {\it Solid line}:
power-law fit $N(>V_{\rm max})=(1/48) \times(V_{\rm max}/V_{\rm max,host})^{-3}$.
{\it Solid lines with error bars:} observed number of dwarf galaxy satellites around the Milky Way
(as in \protect\citet{Mateo1998,Klypin1999}, plus the recently discovered Ursa Major dwarf 
\protect\citep{Willman2005,Kleyna2005}),
within $\rvir$ and within $0.1\,\rvir$ (only Sagittarius).
The vertical error bars show the Poissonian $\sqrt{N}$ scatter.
}
\vspace*{0.1in}
\label{velf}
\end{figure*}

Figure \ref{velf} shows the cumulative peak circular velocity 
($V_{\rm max} \equiv {\rm max} \{ \sqrt{GM_{\rm sub}(<r)/r} \}$) function for the
entire subhalo population within $\rvir$ as well as for the inner sub-population.
The former is well-fit by a power-law 
$N(>V_{\rm max})=(1/48) (V_{\rm max})/ V_{\rm max,host}) ^{-3}$ 
above $5\,\kms$, as shown by the dotted line.
This is in good agreement with the mean abundance obtained from a large sample of 
host halos by \citet{Reed2005} after accounting for the effect of the
different cosmology.\footnote{\citet{Reed2005} simulations were
normalized using $\sigma_8=1.0$. Decreasing $\sigma_8$ to 0.74 (the value used here) 
lowers the subhalo velocity function by about a factor two (see \citealt{Zentner2003} 
and the erratum to \citealt{Diemand2004sub}).}
Note how numerical effects start to flatten the velocity functions
below $5\,\kms$ and already below $15\,\kms$ for the lower resolution
run.

\subsection{A "missing inner satellites" problem?}
\label{satellites}

The number of dark matter subhalos with peak circular velocity
larger than $10\,\kms$ is 124 (compared to 74 in the lower resolution
run). This is an order of magnitude larger than the dozen or so dwarf
satellite galaxies of comparable circular velocity observed around the
Milky Way \citep{Moore1999,Klypin1999}, a well known mismatch often
referred to as the ``missing satellites problem".  The {\it WMAP}
3-year cosmological parameters do not alleviate the discrepancy
significantly, and with our very high resolution Via Lactea the difference
between the predicted number of CDM subhalos and the satellites actually observed has 
become even more pronounced.
Indeed, we find a new ``missing inner satellites problem" within $0.1\,\rvir=39\,$kpc, a
region that appeared practically smooth in previous lower resolution
simulations. In this inner region we now resolve 5 (26) dark satellites
with $V_{\rm max}>10\,\kms$ ($V_{\rm max}>5\,\kms$), compared to only
one known Milky Way dwarf galaxy, Sagittarius at a distance of $24\pm
2$ kpc \citep{Mateo1998}.  Even when the suspected stripped cores of
dwarf spheroidals from the recent compilation of
\cite{vandenBergh2006} are included, the numbers of observed inner
satellites remain well below the predicted number of substructures in the inner halo.

Recent Local Group models \citep{Kravtsov2004,Moore2006} were based on simulations
like the lower resolution run used in this work. The overall abundance of subhalos does
increase with resolution in the relevant $V_{\rm max}$ range, but not dramatically, 
i.e. the total abundance of satellite galaxies in these models seems not to be very 
affected by their lower numerical resolution.
But what would these two models predict about satellite galaxies within the newly resolved
inner subhalos? The \cite{Kravtsov2004} model allows galaxy formation only in subhalos
that were relatively {\it massive} ($> 10^9\,\msun$) before they fell into the gravitational 
potential of the host. In the \citet{Moore2006} scenario only the {\it earliest} forming 
halos above the atomic cooling mass at $z>12$
become proto-galactic building blocks (many of these systems remain well below 
$10^9\,\msun$ at all times). By $z=0$ they have built up a halo of stars, globular clusters and a 
few surviving dwarfs. We followed the assembly history
of the inner subhalos backwards in time and found that only two of them
lie above the (time dependent) minimum mass from \cite{Kravtsov2004}.
The same two systems are also the only ones that form early enough to become luminous in the
\citep{Moore2006} scenario. In the inner $0.1\,\rvir$ both models thus predict
two dwarf galaxies, in good agreement with the numbers observed around the Milky Way and M31,
and leave the other three inner subhalos above $10\, \kms$ dark. Therefore the new, missing
inner satellites problem seems to be resolvable in the same way as the well known Local Group wide
problem. The overabundance of inner subhalos might not be a problem for CDM, but it leads to
the new and interesting prediction of having a number of relatively large, dark CDM subhalos 
orbiting in the inner halo, i.e. in the same region where the Milky Way galaxy is located.

\subsection{Mass fractions in substructure and gravitational lensing}

\begin{figure*}
\plotone{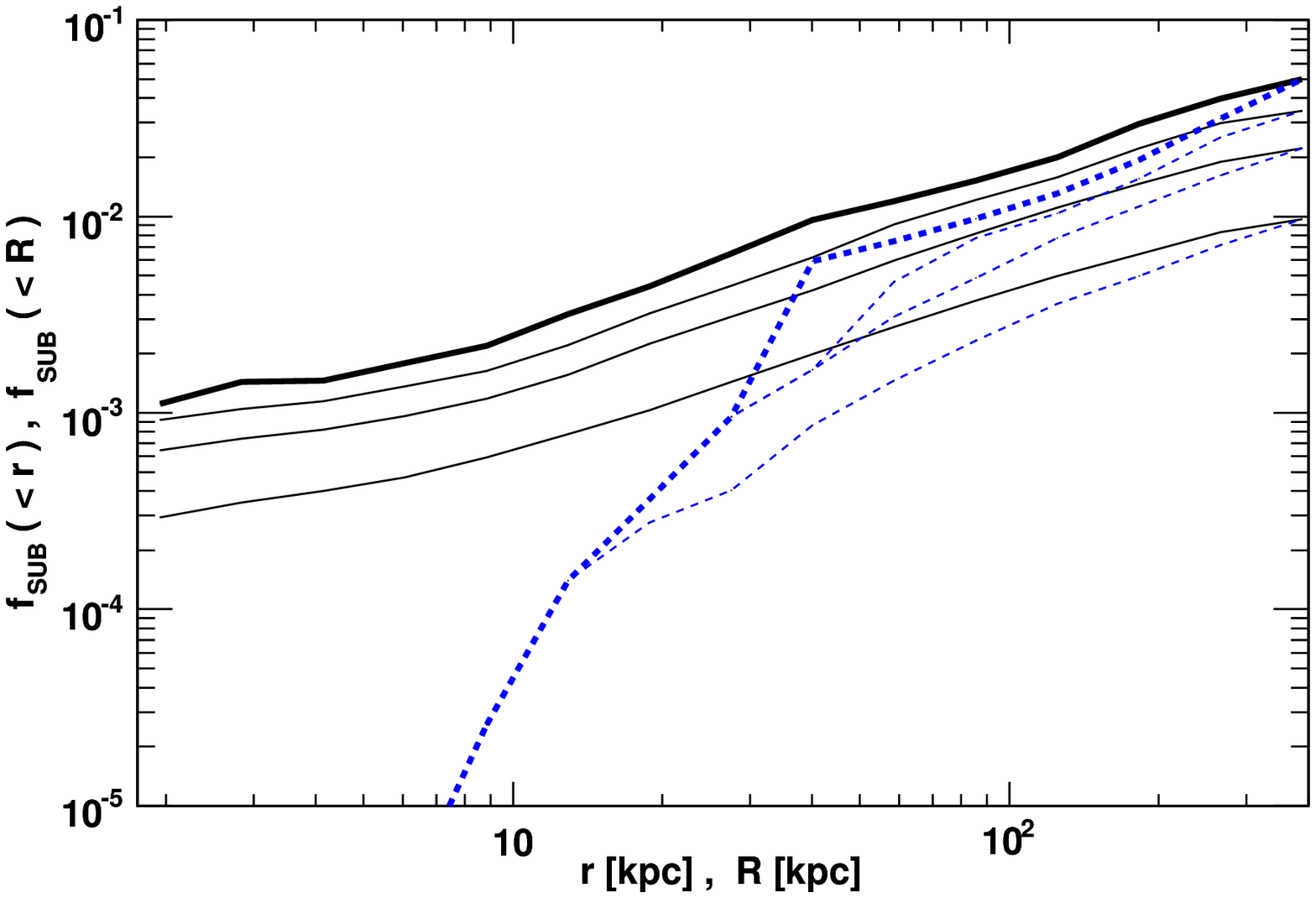}
\caption{Cumulative mass fraction of substructure, shown spherically-averaged as a function
of radius $r$ ({\it dashed lines}) and in cylindrical projection as a function of
projected radius $R$ ({\it solid lines}). Averages over 
1000 random projections are shown, and only subtructures within $\rvir$ are included. 
{\it Thick lines:} all subhalos. {\it Thin lines:} subhalos 
with $\msub<(10^7\,\msun,10^8\,\msun,10^9\,\msun)$ from bottom to top. 
}
\vspace*{0.1in}
\label{surf}
\end{figure*}
Standard smooth gravitational lens models have difficulties in explaining the relative
fluxes of multiply-imaged quasars \citep{Mao1998,Metcalf2002}. The discrepancy between the 
predicted and observed flux ratios is commonly referred to as the ``anomalous flux 
ratio problem'', and dark matter substructure within the lens halo is one of the 
leading interpretations of such anomaly (e.g. \citealt*{Metcalf2001,Chiba2002,
Dalal2002,Bradac2004,Mao2004,Amara2006,Maccio2006}).
Rather than the total mass fraction, lensing 
observations are sensitive to the mass fraction in substructure projected through a cylinder 
of radius 5-10 kpc around the lens center. Figure \ref{surf} depicts the fraction of the 
host halo mass within a sphere of radius $r$ that is bound up in substructure,
$f_{\rm sub}(<r)$, as well as the substructure mass fraction in cylindrical projection 
of radius $R$, $f_{\rm sub}(<R)$, measured in our Via Lactea simulation.  The radial distribution follows 
the subhalo number density profile given in \cite{Diemand2004sub},
i.e. it is more extended than the overall mass distribution.
In the vicinity of $R=10\,$kpc, the surface mass density for all subhalos with $\msub<10^9\,\msun$ 
can be approximated as $f_{\rm sub}(<R)=0.002\,(R/10\,{\rm kpc})$. 

Our total projected surface densities are consistent with, but on the low 
side of, estimates from semianalytic models \citep{Zentner2003}, although
we find significantly larger contributions from the smallest subhalos at large
projected radii. The total projected surface densities is lower
than the few percent value that seems to be required to explain the anomalous flux ratios
\citep{Dalal2002,Metcalf2004}. Whether this constitutes a serious disagreement with
CDM predictions is unclear. Here we just notice that our simulation likely suffers 
the numerical ``over-merging'' problem below $4\times 10^6\,\msun$: if the mass spectrum 
of substructure we measure, $dN/d\msub\propto \msub^{-2}$, extended all the way down to 
$10^4\,\msun$, for example, then the mass fraction in 
subhalos with $10^4\,\msun<\msub<4\times 10^6\,\msun$ would be comparable
to that in all higher mass satellites with $4\times 10^6<\msub<10^9\,\msun$ identified in our run.
We also have to expect some ``over-merging'' in the dense, inner halo: Via Lactea 
is able to resolve a few subhalos within 10 kpc of the galactic center (3 systems at $z=0$),
but the true subhalo abundance might be significantly higher in this region. 

\section{Dark matter annihilation signal}
\label{sec:annihilation}

Our simulation shows that, in the standard CDM paradigm, galactic halos should be filled with 
at least tens of thousands subhalos that have no optically luminous counterpart. If the dark 
matter is in the form of a supersymmetric particle produced in the early Universe like the neutralino, 
then substructure may be lit up by the annihilation of such particles into $\gamma$-rays.
Since the annihilation rate is proportional to density squared, the predicted flux 
depends sensitively on the clumpiness of the mass distribution.
\footnote{Annihilation is too slow to alter the mass distribution on
the scales simulated here. For an NFW subhalo it would erase the cusp
only on scales smaller than a micro-parsec and this would not lower the
total subhalo luminosity significantly.}
Within the Milky Way, nearby subhalos
may be among the brightest sources of annihilation radiation, and could be 
detectable by the forthcoming {\it Gamma-Ray Large Area Telescope} ({\it GLAST}).   

\begin{figure*}
*\epsscale{1.3}
\plotone{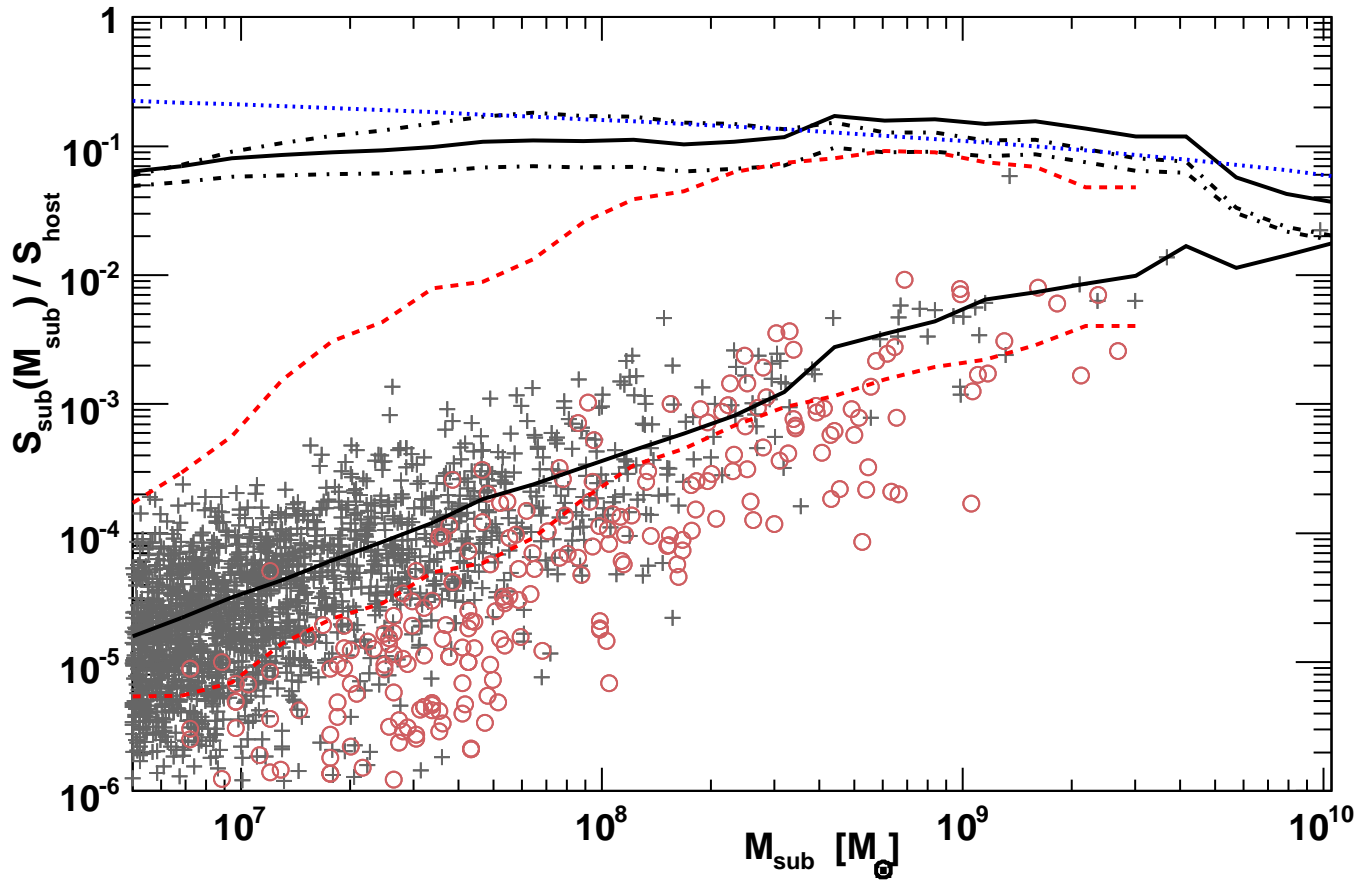}
\caption{The annihilation signal of individual subhalos in units of 
the total luminosity of the spherically averaged MW host.
{\it Crosses}: Via Lactea run. {\it Circles}: lower resolution run. The plotted range
begins at a subhalo mass corresponding to 240 particles in Via Lactea 
and only 9 particles in the lower resolution run.
Densities are estimated 
using the SPH-kernel over 32 nearest neighbors. {\it Solid lines}: 
sliding average ({\it lower curve}) and total signal over one decade in mass ({\it
upper curve} for the Via Lactea run. 
{\it Dashed lines}: same for the lower resolution run. 
{\it Dash-dotted lines}: Alternative estimate of the total subhalo signal
in the Via Lactea run based on densities measured in spherical bins around
the subhalo centers ({\it lower curve}), and same ({\it uppper curve})
based on NFW fits to these density profiles.
{\it Dotted line:} mass dependence of the total signal from
\protect\cite{Colafrancesco2005}, see text for details. 
}
\vspace*{0.1in}
\label{fig3}
\end{figure*}

Annihilation luminosities based on densities measured directly in 
collisionless N-body simulations
\footnote{In this work we neglect the still somewhat 
unclear effects of galaxy formation on the inner dark matter
density. Substantial gains are possible, e.g. the contracted NFW halo "NFW$_c$" in
\citet{Mambrini2006}, rescaled to match Via Lactea,
gives an eight times larger host halo signal than the
corresponding NFW profile before contraction.}
suggest that bound substructure may boost 
the signal from individual halos by only a small factor compared to a smooth 
spherical density profile \citep{Stoehr2003,Diemand2006}, and that the enhancenement
is dominated by the most massive subhalos. The measured contribution of resolved 
substructure must be affected by numerical resolution, however, since higher
resolution simulations would be able to resolve higher densities in subhalo 
centers \citep{Kazantzidis2004}. Indeed, analytic calculations
tend to find larger substructure boost factors \citep{CalcaneoRoldan2000} 
and a signal that is dominated by small subhalos instead \citep{Colafrancesco2005}.
In Figure \ref{fig3} we show the mass dependence of the annihilation luminosity
for individual subhalos in our two Milky Way simulations. The signal from the $i^{\rm th}$ 
subhalo is proportional to
\be
S_i = \int_{V_i} \rho^2_{\rm sub} dV_i = \sum_{j \epsilon \{P_i\}} \rho_j m_p,
\label{eq:annihilation}
\ee
where $\rho_j$ is the density of the $j^{\rm th}$
particle, and $\{P_i\}$ is the set of all particles belonging to halo
$i$. In the Via Lactea run we find $S_{\rm sub}\propto M_{\rm sub}$,
i.e. a signal-to-mass ratio that is approximately independent of mass. 
Given our substructure abundance of 
$dN/d\msub\propto \msub^{-2}$, this implies an annihilation luminosity that
is approximately constant per decade of substructure mass, as the Figure shows.
A comparison with our lower resolution simulation indicates that more than 500 dark matter
particle per subhalo are needed to produce a signal that is
of the right order of magnitude. Below this scale (i.e. below about
$10^7\,\msun$ in Via Lactea and $3\times10^8\,\msun$ for the lower resolution run)
the luminosity decreases quickly due to insufficient resolution.

To test the robustness of the above results we have tried two alternative ways to compute 
matter densities and annihilation luminosites besides the standard SPH-kernel. 
The first is based on densities measured in spherical bins around each subhalo,
while the second uses NFW fits to these density profiles.
All three methods agree with an approximately constant total signal per
decade of substructure mass above a mass scale of 500 particles
per subhalo. However, the spherical density profile
estimate falls below the NFW estimate even for massive subhalos and the 
difference increases with decreasing subhalo mass. This is consistent with the
result of \citet{Kazantzidis2004} that well resolved subhalos
remain as cuspy as halos in the field, while under-resolved structures have relatively large, 
artificial constant density cores. Annihilation luminosities based on local SPH densities,
on the other hand, are larger than the spherical signal for massive subhalos because
of resolved sub-substructures, triaxiality, and other inhomogenities that get
averaged out in the spherical estimates. Smaller subhalos, containing a few thousand particles,
appear rather smooth and spherical and have a large artificial core. 
It is therefore not surprising that in this range the SPH estimate lies above the
spherical density profile estimate, but below the NFW signals.
Below about 3000 particles ($6\times10^7\,\msun $) the NFW total signal
estimate decreases and approaches the other two estimates, presumably because
it too starts to be affected by the finite resolution: with this low number of particles,
halo density profiles are artificially flattened
by numerical effects out to a radius of about 
$N^{-1/3} \,r_t= 0.07 \,r_t$ \citep{Diemand2004cluster}.
The scale radii of many subhalos would lie in the affected range
and the resulting apparent scale radii of the NFW fits come out too large.
This leads to underestimated scale densities and $\gamma$-ray luminosities.
Above 3000 particles, the total signal estimate based on NFW fits hints at a contribution
to the total signal that increases with decreasing subhalo mass
in agreement with the analytic model by \cite{Colafrancesco2005}.

At our highest resolution we measure a total neutralino $\gamma$-ray
luminosity from the host halo that is a factor of 2 higher than the
total spherically-averaged smooth signal: $41\%$ of this increase is
associated with substructure, the other 59\% is due to other
deviations from spherical symmetry.  At ten times lower resolution
\citet{Stoehr2003} found an increase from subhalos of 25\%, and in our
27 times lower resolution run we measure only a 12\% increase.  Since
the subhalo signal appears to be dominated by the smaller clumps, both
the total and the subhalo signal have not converged yet and might be
much larger for a real $\Lambda$CDM halo: assuming a constant boost
factor of 0.17 per decade in subhalo mass from
$10^{-2}\,\msun$\footnote{Micro-subhalos below $10^{-2}\,\msun$ appear
to contribute little to the annihilation signal of their host
\citep{Diemand2006}.}\ to $10^{10}\,\msun$ would give a total
enhancement factor of 3, i.e. the substructure signal would be twice
that of a smooth spherical host halo. Allowing the same boost for each
of the subhalos (0.17 per decade from $10^{-2}\,\msun$ to
$0.01\,\msub$) in a self-similar manner, the total boost for a galaxy
halo is approximately 13. A contribution that grows with decreasing
subhalo mass is not excluded: \cite{Colafrancesco2005} found such a
behavior using an analytic model with a $dN/d\msub\propto
\msub^{-1.9}$ subhalo mass function, subhalo concentrations
proportional to those given for field halos in \citet{Bullock2001},
and somewhat different cosmological parameters ($\Omega_M = 0.281$,
$\Omega_\Lambda = 0.719$, $n=1.0$, and $\sigma_8=0.89$).  Figure
\ref{fig3} shows that the results from the high resolution run are
consistent with their mass dependence, especially after considering
that numerical effects tend to reduce the substructure signal and that
their importance is gradually growing with decreasing subhalo mass.
Integrating over the \citet{Colafrancesco2005} model normalized to our
result at large subhalo masses gives a boost factor of about 8. The
increase due to sub-substructures adds another factor of about ten, it
is more dramatic than in the constant S/M case.  We plan to constrain
this factor further by using analytic models based on the detailed
subhalo and sub-subhalo properties in a future
paper.  

\section{$\gamma$-rays from MW substructure}

\begin{figure*}[htp]
\centerline{\epsfig{figure=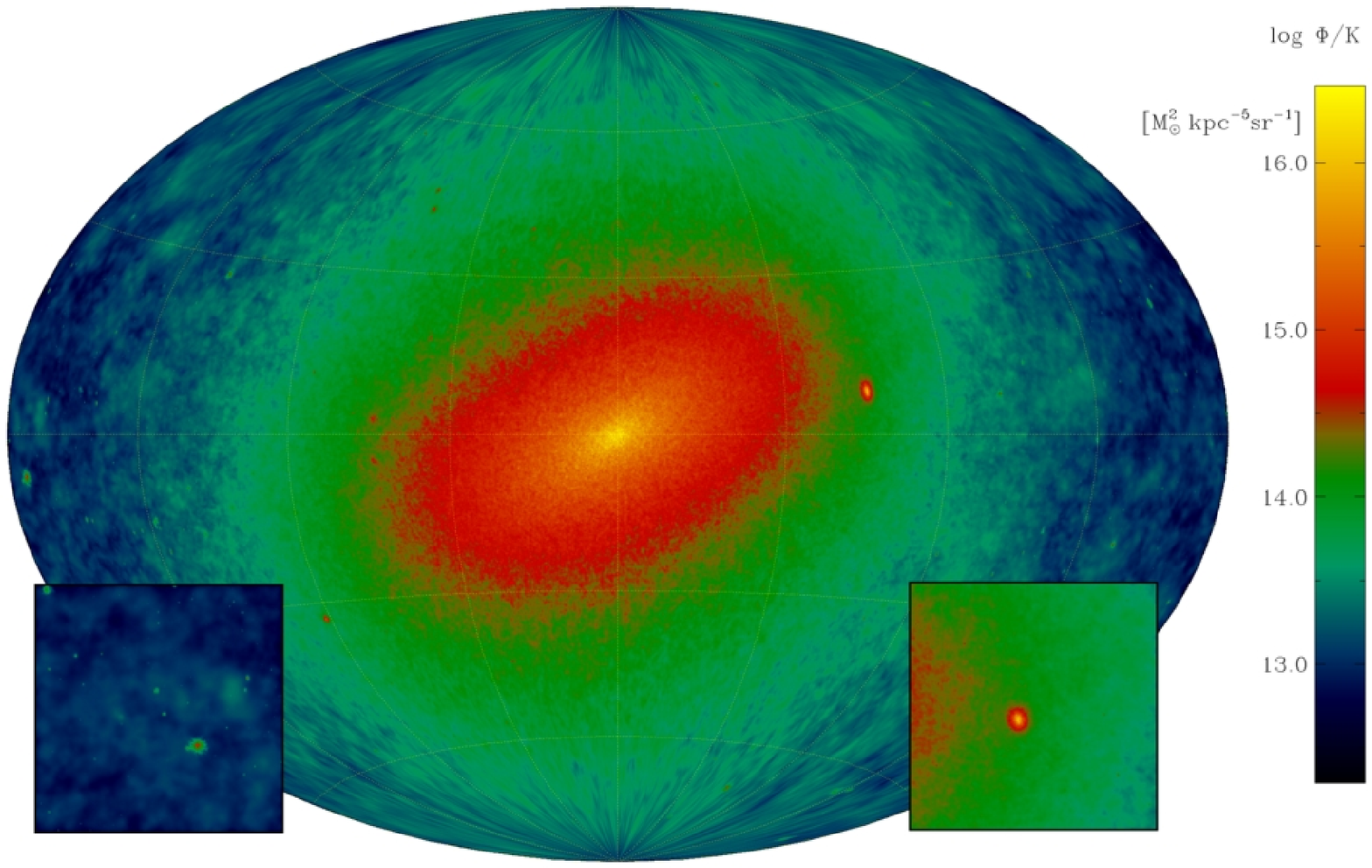,width=7.in}}  
\vspace{0.5cm}
\caption{
All-sky map of the DM annihilation flux $\Phi/K$ ($\msun^2 {\rm
kpc}^{-5}$ sr$^{-1}$) in our Via Lactea halo, for an observer located 8 kpc from 
the Galactic center. The insets show zoom-in's of a $40^\circ
\times 40^\circ$ region around the anticenter (\textit{left}) and the
brightest subhalo (\textit{right}).
}
\label{fig:allsky}
\end{figure*}

The Large Area Telescope (LAT) aboard the soon-to-be-launched {\it GLAST}
satellite may be able to detect $\gamma$-rays from DM annihilations
originating at the Galactic center (e.g. \citealt{Mambrini2006}), in a subhalo 
(e.g. \citealt{Koushiappas2004}) or even in a very nearby micro-subhalo 
\citep{Koushiappas2006}.
The LAT has an effective area of more than 8000 cm$^2$, a field of view greater
than 2 steradian, and sub-degree angular resolution at energies
greater than 1 GeV. {\it GLAST} is expected to operate for more than 5
years, and will conduct an all-sky survey during this time.

We have constructed all-sky maps of the expected annihilation
$\gamma$-ray flux reaching a fiducial observer located 8 kpc from the
center of the Milky Way halo. Each DM particle was assigned an
annihilation $\gamma$-ray flux
\be
F_i = K \f{\rho_i m_p}{4 \pi d_i^2}, 
\ee
where $d_i$ is the distance from the observer to the $i^{\rm th}$
particle. The constant of proportionality $K$ captures the uncertain
particle physics and is equal to $N_{\gamma} \langle \sigma v
\rangle_0/(2 m_\chi^2)$, where $N_\gamma$ is the number of photons
produced per annihilation, $\langle \sigma v \rangle_0$ is the
annihilation rate at zero temperature, and $m_\chi$ the restmass of
the DM particle. In this work we consider only the dependence of the
signal on the macroscopic DM density distribution, and present fluxes
in units of $K$, but for any given particle physics model our results can
be converted to physical fluxes by multiplying by the appropriate
$K$. We also calculate the angular scale subtended by each DM
particle, $\Delta\theta_i = h_i / d_i$, where $h_i$ is defined to be
half the radius that encloses the nearest 32 neighbors. We then bin up
each particle's contribution in angular bins (pixels) of size
$\Delta\Omega$, typically chosen to be $0.1^\circ \times 0.1^\circ
\approx 3\times10^{-6}$ sr.  Each particle contributes to all pixels
within its angular scale, weighted by a projected SPH smoothing
kernel. The weights are normalized such that the sum over all covered
pixels equals the particle's total flux $F_i$. After binning all
particles, we divide each pixel by $\Delta\Omega$, resulting in a map
of flux per solid angle $\Phi$. This 2D image is mapped onto a
Hammer-Aitoff equal area projection, with the coordinate system
rotated such that the halo center lies at the center of the
projection.

The image of Figure~\ref{fig:allsky} depicts the resulting DM
annihilation all-sky map. As expected the signal is strongest towards
the center, with a flux about four orders of magnitude larger than the
diffuse contribution from the anti-center. Several substructures are
clearly visible, both towards the halo center and in the opposite
direction (towards the edges of the image). The two insets show in
greater detail a $40^\circ \times 40^\circ$ region around the
anti-center and around the brightest subhalo. Because the diffuse
background noise is strongest near the center due to Galactic
continuum emission \citep{Stoehr2003}, it is likely that substructure
will be more readily detectable at anti-center, and/or at higher
latitudes.

\begin{figure*}
\plotone{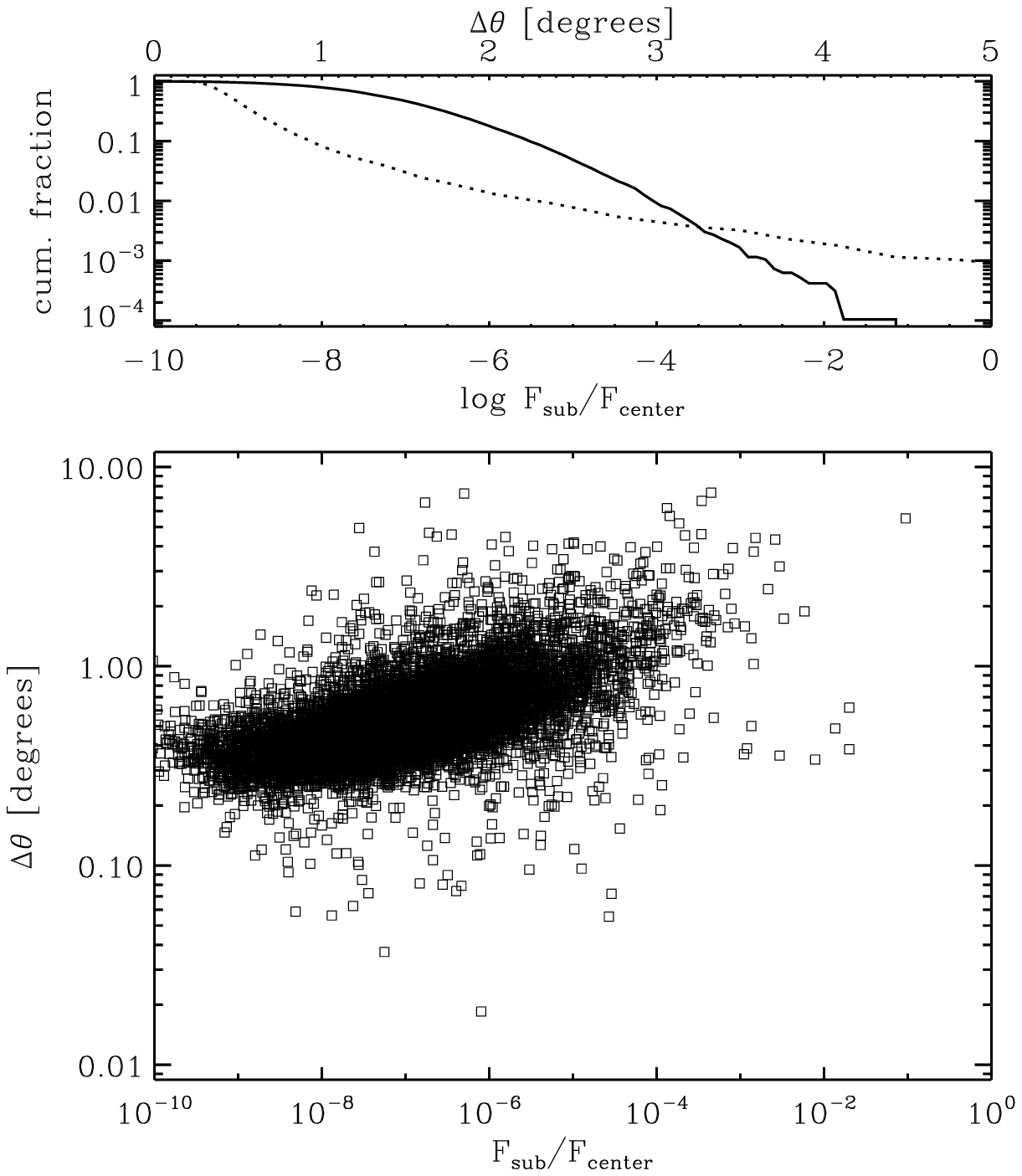}
\vspace{0.5cm}
\caption{
\textit{Top}: Cumulative distribution functions of the ratio of
subhalo flux to central flux $F_{\rm sub}/F_{\rm center}$
(\textit{solid line}) and of subhalo angular size $\Delta\theta$
(\textit{dotted line}), for one random observer position (the same one 
as in Figure~\ref{fig:allsky}). \textit{Bottom}: $\Delta\theta$ plotted
against $F_{\rm sub}/F_{\rm center}$ for all subhalos. Only subhalos
within $\rvir$ were used in both plots.
}
\label{fig:subhalo_size_vs_flux}
\end{figure*}

Compared with the density squared projection in Figure~\ref{L200}
however, it is striking how little substructure is actually visible in
the all-sky map. This is due to a number of causes. At 8 kpc ($\approx
0.02\,\rvir$) the observer is located relatively close to the halo
center. The mean density at this distance is $6.0\times10^4 \rho_{\rm crit}$,
and only 4009 of our subhalos have central densities greater than
this. Of course this is largely a numerical effect, since subhalos are
resolved with much fewer particles than the host. Assuming a universal
density profile and higher subhalo concentrations, subhalos would be
denser than their host halo at comparable $r/\rvir$. Moreover
Via Lactea still resolves only 32 subhalos
closer than 20 kpc from the observer. In our all-sky map distant
subhalos are hard to detect because both their fluxes and angular
sizes on the sky are reduced. In Figure~\ref{fig:subhalo_size_vs_flux}
we have plotted cumulative distribution functions of $\Delta\theta$
and $\log(F_{\rm sub}/F_{\rm center})$, as well as a scatter plot of
the two quantities against each other. Here $\Delta\theta$ is the projected
angular size of the subhalo's tidal radius, $F_{\rm sub}$ is the total
flux from all particles within each subhalo's tidal radius, and
$F_{\rm center} = 4.7\times10^{13} \msun^2 {\rm kpc}^{-5}$ is the flux
from all pixels within $1^\circ$ of the Galactic center. We have only used
subhalos within the host halo's $\rvir$. The median angular size of
all subhalos is $0.48^\circ$, and only 5 subhalos have a flux greater
than $1\%$ of the halo center. The brightest subhalo, at (longitude,
latitude) $= (\phi,\theta) \approx (60^\circ, 10^\circ)$, has $F_{\rm
sub}/F_{\rm center}=0.094$. This subhalo has a mass of $\sim
1.3\times10^{9}\,\msun$, a peak circular velocity of $40\,\kms$  and lies
at a distance of 31 kpc from the Galactic center. It is very unlikely that
our Milky Way halo would contain such a massive DM subhalo so close to
the center, since it would presumably host a dwarf galaxy 
\footnote{This subhalo would host a dwarf galaxy according to both models
discussed in Section \ref{satellites}.}.
The closest such object in the Local Group is the Small Magellanic Cloud, it has a
comparable mass and velocity dispersion \citep{Harris2006}, but lies at
a distance of $\approx 50$ kpc. It would thus appear 1.7 times smaller
and 2.8 times fainter than the brightest subhalo in Figure~\ref{fig:allsky}.

\section{Conclusions}

We have reported in this paper first results from the highest
resolution simulation to date of Galactic DM substructure. Our
simulation consists of over 200 million DM particles and follows the
evolution and formation of a Milky Way-size halo with $\mvir=1.8\times
10^{12}\,\msun$ in a {\it WMAP} 3-year cosmology. Here we summarize our main
results:

\begin{itemize}

\item At the present epoch we resolve approximately 10,000 subhalos,
about one order of magnitude more than in any previous simulation of a
Milky Way halo. Our resolution is sufficient to even resolve a few
subhalos within the solar circle. In several of our more massive
subhalos we are able to identify sub-substructure.

\item The cumulative subhalo mass function is consistent with a
$(\msub/\mvir)^{-1}$ power law down to $\msub=4\times 10^6\,\msun$
(200 particles), implying an equal mass contribution per decade of
subhalo mass. As such the total mass fraction in substructure is going
to be resolution dependent, and our value of 5.3\% is likely to be a
lower limit. Extrapolating the $dN/d\msub\propto \msub^{-2}$ subhalo
mass function down to $1\msun$, for example, would lead to a
substructure mass fraction of 20\%. The fraction of surface mass
density in resolved substructure within a projected distance of 10 kpc
is 0.3\%, again most likely an underestimate.

\item The cumulative subhalo velocity function is also well-fit by a
power law: $N(>V_{\rm max})=(1/48) (V_{\rm max})/ V_{\rm max,host})
^{-3}$ down to $5\,\kms$. We find 124 subhalos with peak circular
velocities greater than $10\,\kms$, about one order of magnitude more
than the number of dwarf satellite galaxies of comparable circular
velocity. The lower $\sigma_8$ and $n$ in the {\it WMAP} 3-year cosmology
do not appreciably alleviate this ``missing satellite
problem''. Furthermore we report a new ``inner missing satellite
problem'' concerning the inner 10\% of $\rvir$. Whereas previous
simulations have found this region to be practically smooth, our
simulation reveals the presence of 5 subhalos with $V_{\rm
max}>10\kms$ and distances less than 39 kpc, compared to only one
known Milky Way satellite, Sagittarius at $24\pm 2$ kpc
\citep{Mateo1998}. A preliminary analysis suggests that this number of
inner satellites is consistent with models in which dwarf galaxy
formation is limited to subhalos sufficiently massive
\citep{Kravtsov2004} or forming in rare density peaks selected at $z>12$
\citep{Moore2006}.

\item The total DM annihilation luminosity of our simulated halo is a
factor of 2 larger than the luminosity derived from a smooth
spherically averaged density profile. 41\% of this increase is due to
resolved substructure and the remaining 59\% can be attributed to
other deviations from spherical symmetry. The luminosity of individual
subhalos is linearly proportional to their mass, which together with
the $\msub^{-2}$ mass function implies an equal contribution to the
total subhalo luminosity from each decade in subhalo mass. Just like
the total subhalo mass fraction, the total subhalo annihilation
luminosity fraction is thus limited by numerical resolution. An
extrapolation down to $10^{-2}\,\msun$ would result in a factor of 3
increase in luminosity. Significantly larger factors due to a
signal per subhalo mass decade which rises slowly with decreasing mass 
and/or similar gains within subhalos due to sub-subhalos cannot be excluded.

\item We have constructed an all-sky map of the expected annihilation
$\gamma$-ray flux reaching a fiducial observer at 8.0 kpc. The signal is
strongest in the center and falls off by about 4 orders of magnitude
towards the anticenter. With our finite numerical resolution only a
small number of subhalos are bright enough to be visible against the
background from the smooth density field surrounding the observer. We
expect actual physical DM subhalos to have higher central densities,
making them more luminous and possibly detectable in regions
sufficiently far from the center. The brightest subhalo in our
simulation has a mass comparable to the SMC, is located 30 kpc from
the center, and has a flux equal to 10\% of the flux from the central
$1^\circ$.

\end{itemize}

\acknowledgments
It is a pleasure to thank Joachim Stadel for making PKDGRAV available 
to us and for writing the code on which our visualization tool is based.
J. D. acknowledges financial support from the Swiss National Science
Foundation and from NASA through Hubble Fellowship grant HST-HF-01194.01
awarded by the Space Telescope Science Institute, which is operated
by the Association of Universities for Research in Astronomy, Inc., for NASA,
under contract NAS 5-26555. P.M. acknowledges 
support by NASA grants NAG5-11513 and
NNG04GK85G, by NSF grant AST-0205738 and 
from the Alexander von Humboldt Foundation. All computations
were performed on NASA's Project Columbia supercomputer system.

{}

\end{document}